**Efficient and Robust Metallic Nanowire Foams for Deep Submicrometer Particulate Filtration**

James Malloy[1,2], Alberto Quintana[1], Christopher J. Jensen,[1] and Kai Liu[1,2,*]

[1]*Department of Physics, Georgetown University, Washington, DC, 20007, United States*
[2]*Department of Physics, University of California, Davis, California 95616, United States*

# Abstract

The on-going COVID-19 pandemic highlights the severe health risks posed by deep submicron sized airborne viruses and particulates in the spread of infectious diseases. There is an urgent need for the development of efficient, durable and reusable filters for this size range. Here we report the realization of efficient particulate filters using nanowire-based low-density metal foams which combine extremely large surface areas with excellent mechanical properties. The metal foams exhibit outstanding filtration efficiencies (>96.6%) in the $PM_{0.3}$ regime, with potentials for further improvement. Their mechanical stability and light weight, chemical and radiation resistance, ease of cleaning and reuse, and recyclability further make such metal foams promising filters for combating COVID-19 and other types of airborne particulates.

**INTRODUCTION**

The ongoing COVID-19 pandemic has unleashed global disruptions and profoundly changed our way of life. Central to the rapid spread of this respiratory infection is the transmission by airborne viral particles.[1-4] The severe acute respiratory syndrome coronavirus 2 (SARS-CoV-2) itself is extremely small, ~ 0.1 μm in size. Often the viral particles are attached to much larger droplets which are released in air in large quantities when an infected person coughs or sneezes. Airborne particles and droplets > 2.5 μm typically precipitate rapidly due to gravity, thus have limited reach. However, particles below this size (labeled $PM_{2.5}$) can suspend in air for hours to days, travel over long distances[5] and spread pathogens such as aerosolized coronaviruses for COVID-19 and the Middle East Respiratory Syndrome (MERS).[6-10] In particular, ultrafine particles below 0.3 μm ($PM_{0.3}$) can penetrate deep into the respiratory system, and some are even able to reach the bloodstream, posing severe health risks.[11-12]

It is critically important to develop efficient, reusable and robust filters for submicron airborne particles, especially in the fight against COVID-19.[1-4] For particles >0.5 μm, they can be captured by filter fibers via interception or inertial impaction.[13] For ultrafine particles (≤0.3 μm) that move around in Brownian motion, trapping occurs due to a concentration gradient of particles between the air and the filter fiber surface, driving small particles towards the fiber surface and binding to them by van der Waals forces.[14] Electrostatic interactions with charged fibers can further aid in the particle filtration, especially at low air flow velocities.[13, 15-17] However, particulates in the intermediate 0.1-0.4 μm size range are the most difficult to filter.[18]

Materials currently used for air filtration have various limitations. For example, filters made of fiberglass fibers are fragile,[19] challenging to clean effectively and use as durable Personal Protective Equipment (PPE),[20] and deteriorate under high temperatures and high relative humidities.[21] Carbon nanotubes (CNT) based filters are prone to mechanical brittleness that leads to CNT release, which may generate new airborne particulates themselves.[18] Polymeric fibers, like Polypropylene fibers that are typically used in high-efficiency particulate air (HEPA) filters and N95 facemasks, are susceptible to degradation upon exposure to UV radiation,[22] or contact to either organic solvents or chlorine-based solutions,[1] making them difficult to decontaminate and reuse. Additionally, polymeric fibers often rely on electrostatics to improve efficiency, which discharge over time, especially under active use, dramatically decreasing the filtration efficiency.[23] The sheer volume of the waste face masks and other PPE materials generated on the global scale amidst COVID-19 also poses significant environmental challenges.[24] On the other hand, recent progresses in metallic foams have presented the possibility of their use as effective filters.[25-28] In addition to the improved structural integrity, these filters are resistant to oils and various organic solvents, corrosive chemicals, ionizing radiation and can survive high temperatures and large pressures without deformation. They can be easily cleaned and decontaminated to allow sustained use, and recycled at the end of useful lifetime.

In this work we report the realization of efficient and durable deep-submicron particulate filters based on low-density metallic nanowire (NW) foams. Such foams are made of randomly distributed Cu nanowires with tunable densities using a scalable electrodeposition and sintering approach. The extremely large surface areas of the foams lead to excellent filtration performance for a broad range of particle sizes, and particularly with efficiencies of > 96.6% for $PM_{0.3}$ particles, similar to N95 facemasks, within breathable pressure differentials (6-60 Pa). Remarkably, this is achieved without the assistance of electrostatics. These light weight metal foams also exhibit outstanding mechanical properties, allowing strong air flows and large pressure drops, and can be easily cleaned, reused and recycled.

# FOAM FILTER SYNTHESIS

Copper foams were fabricated using electrodeposited nanowires following a cross-linking and freeze-drying technique.[27] Anodized aluminum oxide (AAO) templates with 0.2 μm pore size and 60 μm thickness were used for the electrochemical deposition (ED) of Cu NWs.[29-31] Subsequently, NWs were liberated from the AAO membrane into deionized water. The NW-water suspension was then freeze-cast by liquid nitrogen into the desired shape and then pumped in vacuum to sublimate the ice. The resultant free-standing foam was strengthened by sintering and multiple oxidation/reduction cycles.[32] In this work, foams with a density ~ 1% of the Cu bulk density were achieved after this first synthesis step of electrodeposition and sintering (referred to as 1ED-Cu hereafter). Such foams have extremely large surface area-to-volume ratios, up to $10^6$ : 1 $m^{-1}$,[27] that are highly beneficial for filtration. However, these Cu foams were still very brittle, unable to withstand typical air flow through a facemask (about 0.1 m/s),[33] as shown in **Figure S1** of the Supporting Information. These foams were further strengthened with a second ED step (referred to as 2ED-Cu hereafter, more details in the Supporting Information). The final foam density was tuned between 2% and 30% of Cu bulk density.

Typical 2ED-Cu foams plated to 5%-30% bulk density were examined by scanning electron microscopy (SEM). Top-view SEM image (**Figure 1a**) shows that the arbitrary arrangement of interconnected NWs in a 5% foam creates a highly porous structure. Zoomed-in view (**Figure 1b**) reveals that the NW diameter has increased substantially, from the initial 0.2 μm size after the 1$^{st}$ ED to up to 0.5 μm in the foam interior and 0.9 μm on the exterior surface, after the 2$^{nd}$ ED process. Importantly, thickening of the foam not only occurs along the branches but also at the intersections, creating a 3D scaffold over the 1ED-Cu foam where the contact areas between intersecting nanowires are increased by over an order of magnitude. This step is crucial for enhancing the mechanical stability of the foam. SEM image of the 5% 2ED-Cu foam interior reveals that the nanoporous foam morphology is preserved along the sample thickness (**Figure S2**). Moreover, numerous tiny granular textures are observed along the NWs, with size ranging in ~0.1-0.5 μm, resulted from the 2$^{nd}$ ED process (**Figure 1b**). These nucleation/growth sites further increase the overall surface area and fiber surface curvature of the foam. In the 15% density foam, the NW diameter has increased to ~ 1.1 μm in the foam interior and 1.9 μm on the exterior surface, along with reduced surface roughness as the 2ED nucleation sites grow bigger and begin to coalesce (**Figure 1c**). In the 30% density foam, the average NW size has increased to ~5 μm in the foam interior and 6 μm on the exterior surface; the surface roughness is further reduced as the 2ED forms a contiguous coating over the 1ED foam, likely bundling multiple adjacent nanowires into larger μm-sized filaments (**Figure 1d**). Thus with increasing foam density, the surface area per volume decreases substantially.

**MECHANICAL PROPERTIES OF THE FOAM**

The strengthened Cu foams are now mechanically sturdy yet still maintain the low density (~ 2- 30 % of the bulk metals) and light weight. A 120 mg 15% 2ED-Cu foam disc, 1.4 mm thick and 9 mm in diameter, can be easily supported on top of the bristles of a green foxtail plant (Setaria viridis) without bending them (**Figure 2a**). Interestingly, the same foam is able to sustain a heavy load of 1 kg (about 10,000 times its own mass) without collapsing (**Figure 2b**).

To quantify the improvement in strength produced by the electrodeposition process, 2% 1ED-Cu, 2% 2ED-Cu and 15% 2ED-Cu foams are evaluated under compression tests (Supporting Information, **Figure S3**). Clearly the 2ED process has stiffened the foam, since under identical compressive stress the 2ED-Cu foams exhibit smaller deformation (less strain) than the 1ED-Cu foam (**Figure S3a**). The higher stress required to induce the plastic deformation in the 2% 2ED-Cu highlights the improved interwire strength. The 2% 2ED-Cu foam was measured to have a yield strength of 41 kPa, almost an order of magnitude higher than that in the 2% 1ED-Cu foam (5.5 kPa). For the 15% 2ED-Cu sample, no evidence of plastic deformation was observed with a maximum load of 0.6 MPa (6 atm) (**Figure S3b**). Air flow measurement tests performed on 5% 2ED-Cu foams find that they are strong enough to withstand air speeds of over 20 m/s and pressure drops of 0.1 MPa (1 atm) without showing any signs of degradation. These 2ED-Cu foams are thus robust enough to be used as practical filters.

**FILTRATION CHARACTERISTICS**

Pressure differentials across the 2ED-Cu foams, an important parameter to quantify the filter performance, are measured. Small pores will increase the particle filtration efficiency at the expense of breathability, as it will simultaneously increase the pressure drop (PD) across the filter, thus increasing the energy cost to preserve the air flow.[18] Pressure differentials between the up / down-stream have been measured as a function of face velocities, the speed at which air flows through the filter, in the 5-30% 2ED-Cu foams (**Figure 3a**). The measurements confirm the robustness of the foams, successfully withstanding large pressure drops of few tens of kPa at large face velocities of several m/s. PD is found to obey a quadratic relationship with respect to face velocities, $\Delta P = Av + Bv^2$ where $A = (0.97 \pm 0.11)$ kPa m$^{-1}$s and $B = (0.67 \pm 0.01)$ kPa m$^{-2}$s$^2$ for a 1.0 mm thick 15% bulk density foam.[34] A 0.8mm thick 5% foam exhibits an even smaller pressure differential of $A = (0.58 \pm 0.03)$ kPa·m$^{-1}$s and $B = (0.25 \pm 0.01)$ kPa m$^{-2}$s$^2$.

At very low face velocities (< 0.1 m/s), where most common filters operate, this dependence can be approximated to the first linear term.[35] This linear pressure drop coefficient ($A=\Delta P/v$) is inversely proportional to breathability, where a lower value corresponds to a smaller pressure drop and better breathability. The $(0.58 \pm 0.03)$ kPa m$^{-1}$s value measured in the 5% foam indicates high breathability, which is critical for facemask viability, as compared to that of 1.0 kPa m$^{-1}$s for the N-95 respirators and 0.86 kPa m$^{-1}$s for double layered high thread count (HTC) cotton cloth filters.[36] For face velocities of about 0.01-0.1 m/s, PD across our foams are in the 6-60 Pa range, i.e. within the breathable range.[37]

Filtration properties of the synthesized 2ED-Cu foams have been evaluated through two tests. First, a smoke test is performed by burning small white birch wood pellets, generating smoke with $PM_{2.5} = 4\times10^6$ particles·L$^{-1}$, $PM_{10}=2.5\times10^4$ particles·L$^{-1}$ and $PM_{+10}=1.8\times10^3$ particles·L$^{-1}$, and filtered through two consecutive 5% 2ED-Cu foams. While the first foam has suffered a dramatic discoloration due to carbon coating, the second one has barely changed color (**Figure 3b**). Mass change measurements show that most of the smoke particles are captured by the first filter (98% mass), as summarized in **Table S1** (Supporting Information). Moreover, the foam was easily cleaned with 84% of the smoke mass removed after a 3-min rinse in water. Due to the robustness of the foam, sonication and compressed air blowing can be used to further clean the foam, much more desirable than the use of polar organic solvents and high temperature drying processes.[28]

To get a more quantitative measure of the filtration efficiency, a second filtration test was carried out using aerosol-based polydisperse NaCl particles, widely used for testing filters for facemasks.[2-3] These particles are generated by vigorously stirring a solution of 350 g/L NaCl, and they obey a lognormal size distribution.[38] Filtration efficiency is evaluated at 3 particle size ranges, 0.8-1.6 μm, 0.5-0.8 μm and 0.1-0.4 μm, using a set-up illustrated in **Figure 3c**. An aerosol generation chamber is connected to the filter and a vacuum outlet, thus the generated particles flow downstream to the filter. The 2ED-Cu foam is placed in between two polycarbonate (PC) membranes, the first sets the upper size limit of the particle being filtered and the second sets the lower limit, trapping all particles escaping the foam within the selected size range. The mass-based filtering efficiency, $E_m$, is then determined by measuring the total mass of salt trapped in the foam with respect to that collected by the second polycarbonate membrane. Since typical filtration efficiencies are represented in percentage of the number of particles captured ($E_n$), we have converted the mass-based efficiencies ($E_m$) to this conventional unit (more in Supporting Information).[39]

**Table 1** shows the results for the NaCl filtration test for a number of 2ED-Cu foams. Filtration tests reveal that for particles in the 0.8 - 1.6 μm, 0.5 - 0.8 μm, and 0.1 - 0.4 μm size ranges, the mass-based filtration efficiency $E_m$ is 99.4%, 96.1%, and 78.3%, respectively for 1.0 mm thick 15% 2ED-Cu foams, which has benefited from the extremely large surface areas of the foams. For a 0.8 mm thick 5% density 2ED-Cu foam, $E_m$ increases to 85.5% for 0.1 - 0.4 μm sized particles, while the pressure drop coefficient improves to $0.58 \pm 0.03$ kPa m$^{-1}$s. Inertial impaction and interception are the primary mechanisms for filtering the 0.8 - 1.6 μm and 0.5 - 0.8 μm sized particles, respectively. For the critical 0.1 - 0.4 μm sized particles, both diffusion/Brownian motion and interception are effective, at low and high end of the size range, respectively. Improved efficiency in this particle size can be achieved by increasing the foam thickness up to 2.5 mm, reaching a remarkable $E_m$ of 97.0% in a 15% 2ED-Cu foam. The linear pressure drop coefficient increases up to $(2.76 \pm 0.21)$ kPa m$^{-1}$s, still within the breathable regime (see Figure 3a). The converted particle number-based efficiency, $E_n$, is shown in **Figure 3d**. For the hardest-to-filter 0.3 μm particles, $E_n$ is 76.8% and 96.6% for the 1.0 mm and 2.5 mm 15% foam, respectively, and 84.2% for the 0.8 mm 5% foam. Higher efficiencies are achieved for other sized particles in the 0.1-0.4 μm range. For example, for 0.1 μm particles (such as the SARS-CoV-2 virus), $E_n$ of 95.5% and 99.9% is achieved for 1.0 mm and 2.5 mm thick 15% foam, respectively. If the foam density is further increased to 30% of bulk density, the filtration efficiency $E_m$ dramatically drops down to 54.2%, with an increase in $\Delta P/v$ to $(4.67 \pm 0.24)$ kPa m$^{-1}$s (Figure 3a).

It is useful to compare our Cu foam filters with other common filters. As benchmarks we have used our method and setup to test the filtration efficiency of several common filters, including 3M 8000 series N95 respirator (new and washed), 3M 2200 MPR Filter, and high thread count cloth. As shown in **Table S2** (Supporting Information), our measurements are consistent with the ratings of these respective filters. In order to provide an effective gauge for each filter's performance, we introduce a filtration quality factor, $Q \equiv -\frac{v \cdot \ln(\alpha)}{\Delta P}$, where $\alpha$ is the penetration ratio of particles ($\alpha = 1 - E_n$), $v$ is the air velocity (in $m/s$) and $\Delta P$ is the pressure drop across the filter (in kPa). This metric, similar to those in literature,[2-3] is *independent of filter thickness or number of filters*.[40] It is also relatively independent of air flow conditions, as $\Delta P$ linearly depends on velocity in the low velocity regime ($v < 0.1$ m/s) where most filters operate, and consequently $Q$ is reduced to $\sim \ln(\alpha)$. The only $Q$ dependence on air velocity is due to the capture efficiency.

The comparison of the quality factor for 0.3 μm particles $Q_{0.3}$ of the tested filtration media is shown in **Figure 4**. Two layers of 1000 thread count cotton sheets were found to filter 31.3±0.3% ($E_n$) of 0.3 μm average sized particles with a $\Delta P/v$=0.86±0.08 kPa·m$^{-1}$s, which gives a corresponding $Q_{0.3}$=0.44±0.04 kPa$^{-1}$·m·s$^{-1}$. An electrostatically charged 2200 MPR filter was found to have an $E_n$ of 56.2±2.2% and $\Delta P/v$=0.28±0.01 kPa·m$^{-1}$s, resulting in $Q_{0.3}$=2.95±0.10 kPa$^{-1}$·m·s$^{-1}$. An electrostatically charged 3M 8000 series N95 respirator was measured to have an $E_n$ of 96.2±0.2% and $\Delta P/v$=1.02±0.02 kPa·m$^{-1}$s, resulting in $Q_{0.3}$=3.20±0.06 kPa$^{-1}$·m·s$^{-1}$. Upon partial discharging after soaking in water, $E_n$ is reduced to 79.1±0.4%, corresponding to $Q_{0.3}$=1.53±0.03. For our 5% 2ED-Cu foam (0.8 mm thick), with an $E_n$ of 84.2±3.1% and $\Delta P/v$= $(0.58 \pm 0.03)$ kPa·m$^-$

$^1$s, $Q_{0.3}$=(3.18±0.33) kPa$^{-1}$·m·s$^{-1}$. This $Q_{0.3}$ value is comparable to an electrostatically charged N95 respirator, while the breathability, inversely proportional to $\Delta P/v$, is almost twice as good.

## Discussions

The excellent filtration efficiency of the foams is due to the extremely high surface areas of the foams, which enhances the filtration by diffusion, interception, and impaction. This is consistent with previous studies that found thinner fibers filter more efficiently.[28, 40-41] A recent study on 3-dimensional visualization of the filtration process in N95 masks finds a dramatic improvement in capture efficiency for fibers < 1.8 μm in diameter, where the typical fiber diameter ranges in ~ 1-10 μm, with an average size of over 4 μm and an average porosity of 87.3%.[4]

In our 30% density foams, after the 2ED process, the average NW diameter is around 5 μm (Figure 1d), comparable to those in N95 masks. The PM$_{0.3}$ filtration efficiency $E_m$ = 54.2% for a 2 mm thick foam is comparable to the 56% efficiency in *uncharged* N95 masks,[1] without the assistance of electrostatics. In the 15% foams, the surface area has increased significantly, as the NW diameter ranges in 1.1-1.9 μm (Figure 1c). Furthermore, the 2ED process leads to numerous nucleation/growth sites on the surface of the nanowires, significantly increasing the surface area and fiber surface curvature. The foam porosity (85%) is comparable to that in N95 masks, but the surface area is much larger. This correlates with a substantial increase of $E_m$ to 78.3% in a 1 mm thick foam (Table 1). Finally, in the 5% density foams, the nanowire size ranges in 0.5 – 0.9 μm, and the surface is again coated with numerous, even smaller, bumps after the 2ED (Figure 1ab). This foam with the highest surface area : fiber volume ratio, estimated to be over one order of magnitude larger than that in the 30% foam, exhibits the largest $E_m$ = 85.5% for a 0.8 mm thick foam. Thus the lower the foam density, the smaller the NW diameter and the higher the surface area, the higher the filtration efficiency, until the foam mechanical stability becomes compromised in the lowest density foams.

We note that the synthesis of such foam filters is scalable. Using commercially available 1-2" sized nanoporous PC membranes as templates, the materials costs for making nanowires[27, 29, 42] are estimated to be about \$0.35/ml of metal foams, or about \$2 /mask if the active filtration layer is 300 μm thick foam filter media over a 200 cm$^2$ area, the same dimensions as the active filter in an N95 respirator. These costs can be further substantially reduced when the metal foams are produced in industrial scales. Furthermore, the robustness and chemical resistance of the foams allow for easy cleaning and reuse, which will substantially extend the lifetime, and the eventual *cost per use* will be quite competitive with conventional filters and masks.

The foam filters may be used in a wide variety of settings, from respirators/ face masks to household air filters. Flat metal foams can be used in conventional air filter applications without the need for bending, or as inserts in respirators that have filter cartridges. For face masks which require conformal fit, the metal foams may be integrated with elastic coating or back-support, where the PM$_{0.3}$ filtration is accomplished by the metal foams, and the support layer provides the conformal fitting and filtration of larger particulates.[26] Such hybrid structure may also accommodate the application of electrostatics. Additionally, metal foams benefit from antibacterial effects in certain materials.[28] Importantly, metal foams have much better recyclability compared to polymer based filters,[24] minimizing the environmental impact.

## Conclusions

We have demonstrated an effective, durable, reusable and recyclable particulate filter for deep submicron airborne nanoparticles using light weight interconnected metallic nanowire foams. The synthesis method employs freeze-drying and sintering, along with two separate electrodeposition processes, to achieve low density (1-30% of bulk density) metal foams with extremely large surface areas and much enhanced mechanical stability. Such foams can be easily cleaned by rinsing in water, sonication or blowing with compressed air for sustained use. Pressure differential measurements confirm the breathability of the synthesized foams. These Cu metal foams exhibit near 100% filtration efficiencies for 0.1-1.6 μm sized particles, including > 96.6% efficiency for 0.1-0.4 μm particles without the use of electrostatics, and with higher breathability and comparable quality factor as N95 respirators. These results demonstrate a new type of efficient particulate filters, especially for the deep-submicron particulates such as the SARS-CoV-2 virus, that could be used in facemasks, PPE, as well as air/fluid filters in general.

**Supporting Information**.

The Supporting Information is available free of charge on the ACS Publications website at https://pubs.acs.org/doi/10.1021/acs.nanolett.1c00050.

Materials and methods, filtration test on 1ED-Cu foams, additional SEM images of foam morphology evolution, mechanical property measurements, results of the smoke test and conventional filter measurements, and determination of particle number-based filtration efficiency (PDF).

**AUTHOR INFORMATION**

**Corresponding Author**

*Kai Liu - kai.liu@Georgetown.edu

**ORCID**

James Malloy: 0000-0003-2244-1483

Alberto Quintana: 0000-0002-9813-735X

Christopher J. Jensen: 0000-0001-7459-1841

Kai Liu: 0000-0001-9413-6782

**ACKNOWLEDGMENTS**

This work has been supported by the Georgetown Environmental Initiative - Impact Program Award and the McDevitt bequest (Georgetown University), and Tom and Ginny Cahill's Fund for Environmental Physics (U.C. Davis). We are grateful to Profs. Peter Olmsted, Makarand Paranjape, and YuYe J. Tong for helpful discussions, to Dr. Nicholas J. Spada for technical assistance with foam filter holder set up and pressure drop measurements, and to Dr. Xinran Zhang for help with mechanical property measurements.

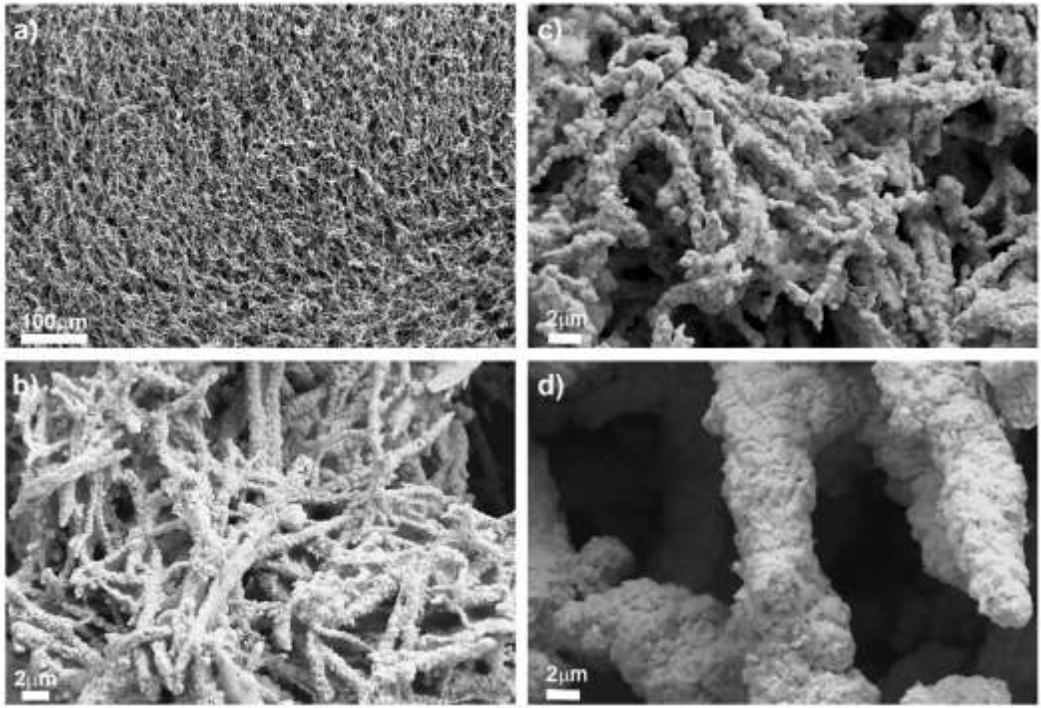

**Figure 1:** SEM images of **a)**, **b)** 5%, **c)** 15%, and **d)** 30% density 2ED-Cu foams, illustrating the evolution of foam morphology and substantial reduction in surface areas with increasing density. **b)** The 5% foam consists of 0.5 – 0.9μm diameter nanowires with numerous tiny bumps whose size ranges in 0.1-0.5 μm. **c)** The 15% foam consists of 1.1-1.9 μm diameter nanowires, and reduced surface roughness as the 2ED nucleation sites start to coalesce. **d)** The 30% foam consists of 5-6 μm sized filaments with the least amount of surface roughness.

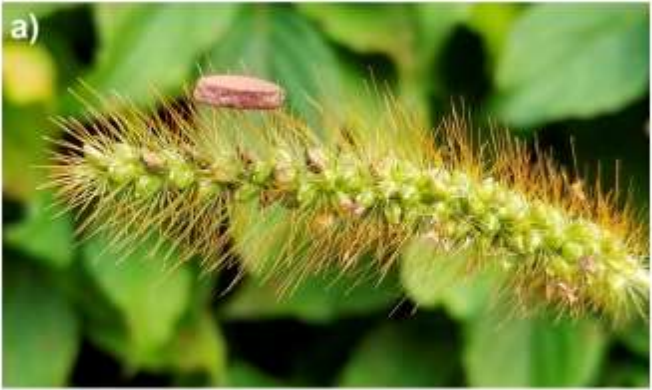
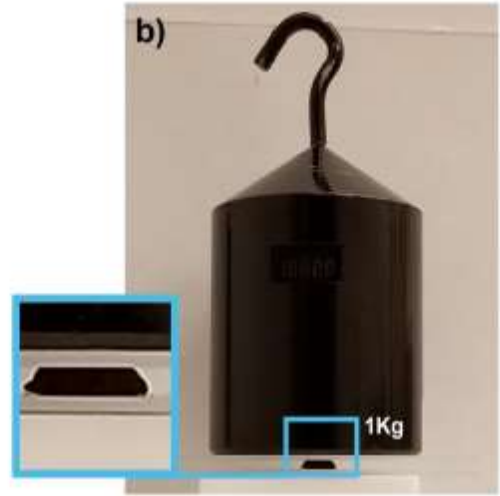

**Figure 2:** A 120 mg 15% 2ED-Cu foam disc, 1.4 mm thick and 9 mm in diameter, **a)** on top of the bristles of a green foxtail plant (Setaria viridis) without bending them, and **b)** supporting a 1 Kg mass, about 10,000 times that of the foam.

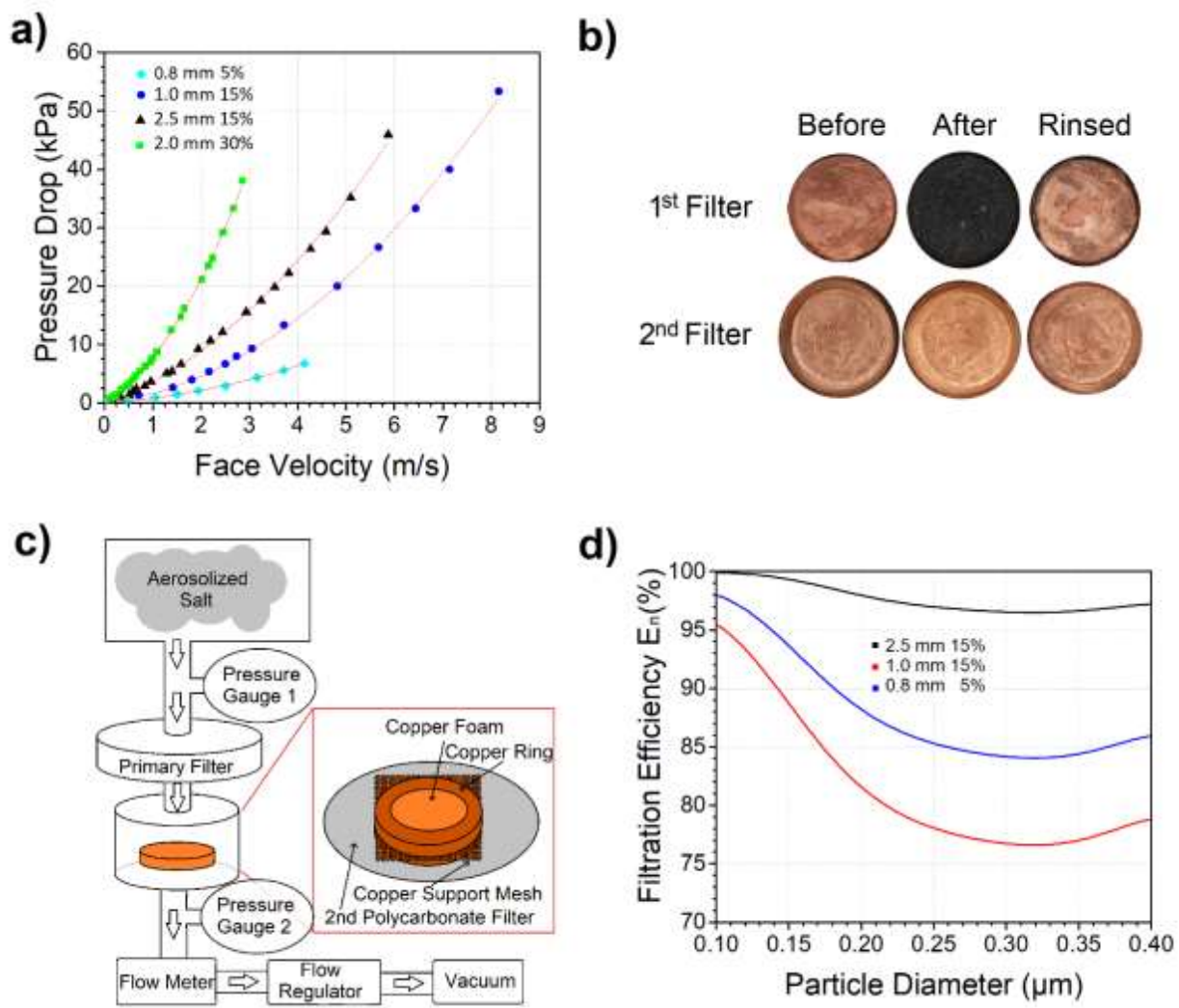

**Figure 3: a)** Pressure drop dependence on the face velocity of 5%, 15% and 30% density 2ED-Cu foams. **b)** Optical images of the two 9 mm sized 5% 2ED-Cu foams before (1st column) and after (2nd column) the smoke test, and after water rinse for a few minutes (3rd column). **c)** Schematic of the experimental set-up used for NaCl particle filtration test. **d)** Extracted filtration efficiency based on number of particles filtered ($E_n$) as a function of particle size for 5% and 15% density 2ED-Cu foams.

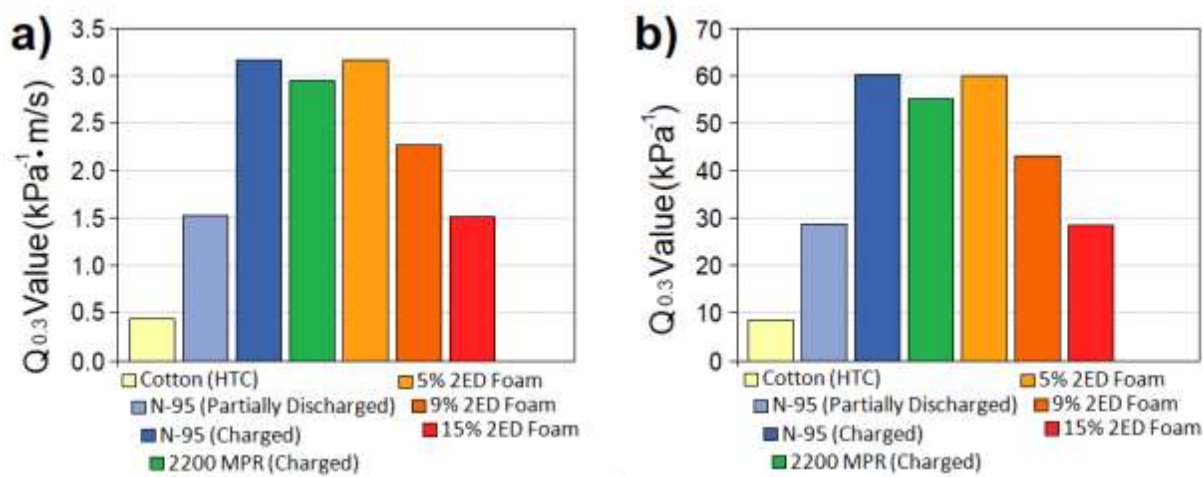

**Figure 4: a)** Comparison of quality factor $Q_{0.3}$ for our synthesized 2ED-Cu foams (0.8 mm thick 5%, 1.0 mm thick 9%, 1.0 mm thick 15%) and other types of filters. **b)** Comparison using the standard quality factor $Q = -ln(\alpha)/\Delta P$ at air velocity $v = 5.3$ cm/s.

Table 1: Summary of 2ED-Cu foam specifications and mass-based filtration efficiency.*

| Foam density (%) | Thickness (mm) | Particle size (μm) | Mass captured by foam (μg) | Mass escaping foam (μg) | Pressure drop coefficient (kPa·m$^{-1}$·s) | Mass capture efficiency $E_m$ (%) |
|---|---|---|---|---|---|---|
| 5 | 0.8 | 0.1-0.4 | 20.5 | 3.5 | 0.58±0.03 | 85.5±3.1 |
| 9 | 1.0 | 0.1-0.4 | 117 | 24.8 | 0.73±0.03 | 82.5±0.5 |
| 15 | 1.0 | 0.8-1.6 | 3300 | 18.0 | 0.97±0.11 | 99.4±0.1 |
| 15 | 1.0 | 0.5-0.8 | 1200 | 47.0 | 0.97±0.11 | 96.1±0.1 |
| 15 | 1.0 | 0.1-0.4 | 47.0 | 13.0 | 0.97±0.11 | 78.3±1.3 |
| 15 | 2.5 | 0.1-0.4 | 50.0 | 1.5 | 2.76±0.21 | 97.0±2.1 |
| 30 | 2.0 | 0.2-0.4 | 67.1 | 56.7 | 4.67±0.24 | 54.2±0.5 |

*Air velocity was between 5-15 cm/s for each foam.

**TOC figure**

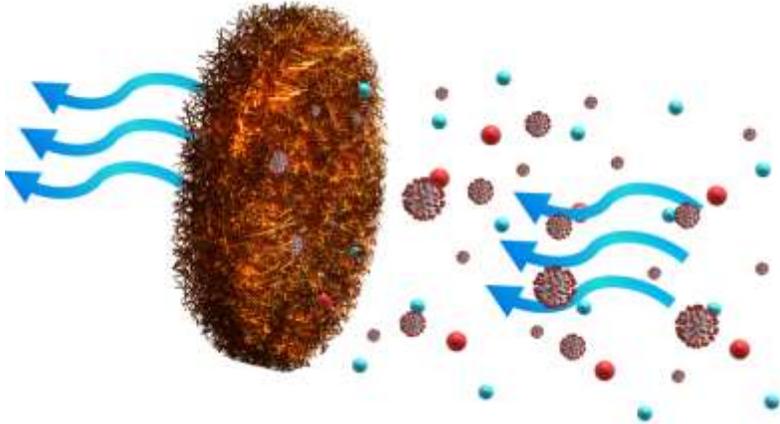